\documentclass[12pt,thmsa]{article}
\usepackage{amsmath}
\usepackage{amsfonts}
\usepackage{amssymb}
\usepackage{graphicx}

\begin{document}
\begin{titlepage}
\begin{center}
{  \bf Random Walk and Diffusion on a Smash Line Algebra }
\vskip
1.0 cm
{\bf Demosthenes Ellinas}\renewcommand{\thefootnote}{*}\footnote{
Email: {\tt ellinas@science.tuc.gr}}
and \
{\bf Ioannis Tsohantjis}\renewcommand{\thefootnote}{**}
\footnote{ Email: {\tt ioannis2@otenet.gr}}
\vskip0.2cm
Department of Sciences\\
Technical University of Crete \\
GR-73 100 Chania Crete Greece\\
\vskip0.2cm
\end{center}
\begin{center}
\begin{abstract}
Working withing the framework of Hopf algebras,  a random walk  and the
associated diffusion
equation are constructed on a space that  is algebraically described as
the merging of the real line algebra with the anyonic line algebra. Technically this
merged structure is
a smash  algebra, namely  an algebra resulting
by a  braided tensoring of  real with anyonic line algebras. The motivation of
introducing the smashing results from the necessity of having non commuting increments
in the random walk.   Based on
the observable-state duality provided by the underlying Hopf structure,
the construction
is cast into two dual forms: one using functionals determined by density probability functions and
the other using the associated Markov transition operator. The ensuing
diffusion equation is shown to possess triangular matrix  realization. The study is
completed by the incorporation of Hamiltonian dynamics in the above
random walk model,  and by  the construction of the dynamical equation obeyed
by statistical moments of the problem for generic entangled density functions.
\end{abstract}
\end{center}
\end{titlepage}

\section{Introduction}

In recent years considerable amount of work has been devoted to generalization
of concepts of random walks and diffusions to spaces described by algebras of
functions or operators, see e.g $\;$the general reviews \cite{schurmanBook}%
,\cite{meyer},\cite{majidbook},\cite{fsbook}. The mathematical framework
mostly adopted for such generalizations is that of Hopf algebras of various
kinds and their braided versions (\cite{abe}, \cite{sweedler}). Remarkably
some of the basic features of random walks and diffusion prosseses, such as
Markov property, convolution, statistical (in)depedence, diffusion as limit of
random walk \cite{majid}, solution of diffusion equations by classes of moment
polynomials (\cite{ffs}, \cite{ellinas}) etc, have their analogue in the
generalized framework. Especially the notion of statistical dependance of
\thinspace steps of a random walk, expressed as the commutation property of
increments of \thinspace walk and the analogue of the central limit theorem
have been extended from the commuting case \cite{GW}, to anticommuting case of
fermionic increments \cite{W}, to the \textit{q}-deformed case (\cite{
schurmanCMP91}, \cite{Lenczewski}), to the case of free products
(\cite{voiculescu}, \cite{speicher}) as well as to case of supealgebras
\cite{asw}. A general approach to independence based on the coproduct of
coalgebras \cite{sLNM86}, can be used to study random walks on braided
structures (\cite{majidplaza}, \cite{franzschott}, \cite{elltso}).

Braiding or smashing defined on algebras, colalgebras, bialgebras and Hopf
algebras, has recently been systematized and a unification of the various
braiding notions has been given (\cite{militaru}, \cite{drabant}). This
compact formulation of smashing products has been used here to intruduce the
so called smash line algebra $\Omega=A\otimes B$, and to study non commutative
random walk and the associated diffusion equation on it. In this present work
we undertake a full investigation of the random walk defined on the space
ensuing from the merging of the real line $A$ with the anyonic line $B$
\cite{majidplaza}. Endowing the algebra of functions of this composite space
$\Omega$ with a smash product, results into smash line algebra. Due to
smashing/braiding, the increments of the walk along the subspace of real line
$A$ are commuting, the increments along the subspace of the anyonic line $B$
are \textit{q}-commuting, while for random steps in the total space $\Omega$
the increments are \textit{Q}-commuting. The outline of the paper has as follows.

In section one, we start with the tensor product algebra of the real line and
the anyonic line algebra and procceed to establish a mathematically
constistent merging of these two in order to form a braided algebra i.e. a
smash algebra. This tensor product algebra is also shown to posses a bialgebra
structure whose $n$-fold coproduct will provides us with the notion of the
$n$-th step of random walk we shall consider. In closing the section we
explain the need of introducing the braiding in that it leads, as it is shown,
to a type of non commutativity among the steps of the random walk. In section
two the diffusion limit of the random walk on the smash line algebra is
obtained and the the associated difussion equation diffusion equationis
studied as well as the moment generating function of the random variable of
the walk. Section three is devoted to the non stationary random walk on the
smash line algebra that expresses the dynamical coupling of the random motion
with a general Hamiltonian dynamics. Towards a physical interpretation of the
constructed random motion on the variables of the smash line algebra, section
four provides a matrix representation of these variables and cast the
diffusion equation in the form of a coupled spin-oscillator type of equation
of motion, for which the solution is obtained in some special cases. Section
five utilizing the state-observable duality built in the Hopf algebra
structure of the random walk and gives a brief construction of the transition
operator of the walk seen it as a markov proccess. In next section six, we
study the effect of the entanglement i.e statistical correlations between the
partial walks of the $x $ variable and the $\xi$ variable, expressed as the
non factorizability of the probability density function ruling the random walk
of the smash line algebra, and investigate, in the terms of the canonical
and\textit{\ q-}deformed ( for \textit{q }root of unity) Heisenberg algebra,
the structure of the ensuing equations of motion for the statistical moments
in the diffusion limit. The last section summarized the rusults and concludes
the paper.

\section{The smash line algebra}

Consider two real associative algebras $\mathcal{A}$ and $\mathcal{B}$ such
that $\mathcal{A}=\mathbf{R}[[x]]$ is the real line algebra, consisting of
formal power series in one variable generated by the element $x$ with
$\{1,x,x^{2},\ldots\}$ as linear basis, and $\mathcal{B}=\mathbf{R}[\xi
]/\xi^{N}$ , the anyonic line algebra consisting of polynomials in one
variable of degree $N-1$, generated by the element $\xi$ with basis
$\{1,\xi,{\xi}^{2},\ldots{\xi}^{N-1}\}$ and ${\xi}^{N}=0$. Both these algebras
are equipped with a Hopf algebra and a smash Hopf algebra structure which will
be examined in turn.

The algebra $\mathcal{A}=\mathcal{A}(\mu_{\mathcal{A}},\Delta_{\mathcal{A}%
},u_{\mathcal{A}},\epsilon_{\mathcal{A}})$ is a commuting, cocommuting, and
coassociative algebra with product, unit, coproduct, and counit respectively
defined as
\begin{align}
\mu_{\mathcal{A}}(x\otimes y)  & =xy,\text{ }\nonumber\\
u_{\mathcal{A}}(c)  & =cI_{\mathcal{A}},\nonumber\\
\Delta_{\mathcal{A}}(x)  & =x\otimes I_{\mathcal{A}}+I_{\mathcal{A}}\otimes
x,\text{ }\label{alga}\\
\varepsilon_{\mathcal{A}}(x)  & =0,\text{ and }\varepsilon_{\mathcal{A}%
}(I_{\mathcal{A}})=1\nonumber
\end{align}
for all $x$, $y$ in $\mathcal{A}$, $c\in\mathbf{R}$ and $I_{\mathcal{A}}$,
being the identity element. Considering the trivial braiding or twist map
$\tau$ on $\mathcal{A}$ such that $\tau(x\otimes y)=y\otimes x$ , $\forall$
$x,$ $y\in\mathcal{A}$, we can introduce a trivial smash Hopf algebra
structure on $\mathcal{A}\otimes\mathcal{A}$, denoted hereafter as
$\mathcal{A}^{\prime}\mathcal{\equiv A}_{\substack{\\\tau}}\bowtie
_{\substack{\\\tau}}\mathcal{A}$\textit{,} \cite{drabant,militaru} with smash
product, unit, smash co-product and co-unit given respectively by%

\begin{align}
\mu_{\mathcal{A}^{\prime}}  & =(\mu_{\mathcal{A}}\otimes\mu_{\mathcal{A}%
})\circ(id_{\mathcal{A}}\otimes\tau\otimes id_{\mathcal{A}})\text{ ,
}\nonumber\\
u_{\mathcal{A}^{\prime}}(1)  & =I_{\mathcal{A}^{\prime}}=I_{\mathcal{A}%
}\otimes I_{\mathcal{A}}\nonumber\\
\Delta_{\mathcal{A}^{\prime}}  & =(id_{\mathcal{A}}\otimes\tau\otimes
id_{\mathcal{A}})\circ(\Delta_{\mathcal{A}}\otimes\Delta_{\mathcal{A}%
})\label{sma}\\
\varepsilon_{\mathcal{A}^{\prime}}  & =\varepsilon_{\mathcal{A}}%
\otimes\varepsilon_{\mathcal{A}}.\nonumber
\end{align}

The smash product $\mu_{\mathcal{A}^{\prime}}$ can easily be extented to an
associative product on the \textit{n}-fold tensor product $\mathcal{A}%
^{\otimes n}\equiv\mathcal{A}^{n}=$ $\mathcal{A}\otimes\mathcal{A}%
\otimes\ldots\otimes\mathcal{A}$ , ($\mathcal{A}$ being taken $n$ times) given by%

\[
\mu_{\mathcal{A}^{n}}=(\mu_{\mathcal{A}^{n-1}}\otimes\mu_{\mathcal{A}}%
)\circ_{k=1}^{2n-3}\circ(id_{\mathcal{A}}^{\otimes^{2n-2-k}}\otimes\tau\otimes
id_{\mathcal{A}}^{\otimes k}).
\]
Then, with $x_{i}=I_{\mathcal{A}}\otimes...\otimes x\otimes_{.}..\otimes
I_{\mathcal{A}}$ , where $x$ is in the i-th position in $\mathcal{A}^{n}$ can
be verified that the braiding relations%

\[
x_{i}x_{j}=x_{j}x_{i}\text{ for all }i,j
\]
are satisfied.

The algebra $\mathcal{B}=\mathcal{B}(\mu_{\mathcal{B}},u_{\mathcal{B}}%
,\Delta_{\mathcal{B}},\varepsilon_{\mathcal{B}})$ is an $N$-potent, commuting,
cocommuting and coassociative algebra with product , coproduct, unit and
counit respectively defined as
\begin{align}
\mu_{\mathcal{B}}(\xi\otimes\eta)  & =\xi\eta,\nonumber\\
u_{\mathcal{B}}(c)  & =cI_{\mathcal{B}},\nonumber\\
\Delta_{\mathcal{B}}(\xi)  & =\xi\otimes I_{\mathcal{B}}+I_{\mathcal{B}%
}\otimes\mathbf{\xi},\nonumber\\
\varepsilon_{\mathcal{B}}(\xi)  & =0,\text{and }\varepsilon_{\mathcal{B}%
}(I_{\mathcal{B}})=1\label{algb}%
\end{align}
for all $\xi$, $\eta$ in $\mathcal{B}$, $c\in\mathbf{R}$ and $I_{\mathcal{B}}%
$, being the identity element. Let us consider this time the braiding or twist
map $\psi$ on $\mathcal{B}$ such that%

\begin{equation}
\psi(\xi\otimes\eta)=q\eta\otimes\xi,\text{ }\forall\text{ }\xi,\eta
\in\mathcal{B}\label{psismall}%
\end{equation}
where $q=e^{2\pi i/N}$, a $N-$th root of unity. We can introduce a smash Hopf
algebra structure on $\mathcal{B}\otimes\mathcal{B}$, denoted hereafter as
$\mathcal{B}^{\prime}\mathcal{\equiv B}_{\substack{\\\psi}}\bowtie
_{\substack{\\\psi}}\mathcal{B}$, \cite{drabant,militaru} with smash product,
unit, smash co-product and co-unit given respectively by%

\begin{align}
\mu_{\mathcal{B}^{\prime}}  & =(\mu_{\mathcal{B}}\otimes\mu_{\mathcal{B}%
})\circ(id_{\mathcal{B}}\otimes\psi\otimes id_{\mathcal{B}})\text{ ,
}\nonumber\\
u_{\mathcal{B}^{\prime}}(1)  & =I_{\mathcal{B}^{\prime}}=I_{\mathcal{B}%
}\otimes I_{\mathcal{B}},\label{smb}\\
\Delta_{\mathcal{B}^{\prime}}  & =(id_{\mathcal{B}}\otimes\psi\otimes
id_{\mathcal{B}})\circ(\Delta_{\mathcal{B}}\otimes\Delta_{\mathcal{B}%
}),\nonumber\\
\varepsilon_{\mathcal{B}^{\prime}}  & =\varepsilon_{\mathcal{B}}%
\otimes\varepsilon_{\mathcal{B}}.\nonumber
\end{align}

The smash product $\mu_{\mathcal{B}^{\prime}}$ can be extented to an
associative product on the n-fold tensor product $\mathcal{B}^{\otimes
n}\equiv\mathcal{B}^{n}=$ $\mathcal{B\otimes B\otimes\ldots\otimes B}$ ,
($\mathcal{B}$ being taken $n$ times) given by%

\[
\mu_{\mathcal{B}^{n}}=(\mu_{\mathcal{B}^{n-1}}\otimes\mu_{\mathcal{B}}%
)\circ_{k=1}^{2n-3}\circ(id_{\mathcal{B}}^{\otimes^{2n-2-k}}\otimes\psi\otimes
id_{\mathcal{B}}^{k}).
\]
Then, with $\xi_{i}=I_{\mathcal{B}}\otimes...\otimes\xi\otimes_{.}..\otimes
I_{\mathcal{B}}$ , where $\xi$ is in the i-th position in $\mathcal{B}^{n}$can
be verified that the braiding relations%

\begin{equation}
{\xi}_{i}{\xi}_{j}=q{\xi}_{j}{\xi}_{i}\text{ for }i>j\;.\label{ij}%
\end{equation}
are satisfied as in \cite{majidplaza}.

The merging of $\mathcal{A}$ and $\mathcal{B}$ as $\Omega\equiv\mathcal{A}%
\otimes\mathcal{B}$ , where $x$ is embedded as $x\otimes I_{\mathcal{A}}$ and
$\xi$ as $I_{\mathcal{B}}\otimes\xi$, admits a bialgebra structure with
product, unit, co-product, co-unit respectively given by%

\begin{align}
\mu_{\Omega} & =(\mu_{\mathcal{A}}\otimes\mu_{\mathcal{B}})\circ
(id_{\mathcal{A}}\otimes\tau\otimes id_{\mathcal{B}}),\nonumber\\
\text{ }u_{\Omega}(1)  & =I_{\Omega}=I_{\mathcal{A}}\otimes I_{\mathcal{B}%
},\label{hopfab}\\
\Delta_{\Omega} & \equiv(id_{\mathcal{A}}\otimes\tau\otimes id_{\mathcal{B}%
})\circ(\Delta_{\mathcal{A}}\otimes\Delta_{\mathcal{B}}),\nonumber\\
\varepsilon_{\Omega} & \equiv\varepsilon_{\mathcal{A}}\otimes\varepsilon
_{\mathcal{B}}.\nonumber
\end{align}
By means of the previous relations and the braiding map $\Psi:\Omega
\otimes\Omega\rightarrow\Omega\otimes\Omega$ defined by%

\begin{equation}
\Psi(x^{^{\alpha}}\otimes\xi^{m}\otimes x^{^{\beta}}\otimes\xi^{n}%
)=q^{mn}Q^{\alpha n+\beta m}(x^{^{\beta}}\otimes\xi^{n}\otimes x^{^{\alpha}%
}\otimes\xi^{m}),\label{braidbig}%
\end{equation}
for all $x,y$ $\in$ $\mathcal{A}$ and and all $\xi$, $\eta$ $\in$
$\mathcal{B}$, $Q\in\mathbf{R}$ and $q=e^{2\pi i/N}$ , we introduce a smash
algebra hereafter denoted by $\Omega^{\prime}=\Omega\#_{\substack{\\\Psi
}}\Omega$. Then $\Psi$ can also be written as
\[
\Psi=(id_{\mathcal{A}}\otimes\Psi_{\mathcal{AB}}\otimes id_{\mathcal{B}}%
)\circ(\Psi_{\mathcal{A}}\otimes\Psi_{\mathcal{B}})\circ(id_{\mathcal{A}%
}\otimes\Psi_{\mathcal{AB}}\otimes id_{B}),
\]
where $id_{A}\otimes\Psi_{AB}\otimes id_{B}$ is realized as%

\[
(id_{\mathcal{A}}\otimes\Psi_{\mathcal{AB}}\otimes id_{\mathcal{B}%
})(x^{^{\alpha}}\otimes\xi^{m}\otimes x^{^{\beta}}\otimes\xi^{n})=Q^{\beta
m}x^{^{\alpha}}\otimes x^{^{\beta}}\otimes\xi^{m}\otimes\xi^{n}
\]
and $\Psi_{A}$, $\Psi_{B}$ are realized as
\begin{align*}
\Psi_{\mathcal{A}}(x^{^{\alpha}}\otimes x^{^{\beta}})  & =x^{^{\beta}}\otimes
x^{^{\alpha}},\\
\Psi_{\mathcal{B}}(\xi^{m}\otimes\xi^{n})  & =q^{mn}\xi^{n}\otimes\xi^{m}.
\end{align*}

\begin{center}%
\begin{center}
\includegraphics[
trim=0.000000pt 533.631775pt 210.884186pt 133.407944pt,
natheight=998.562500pt,
natwidth=705.062500pt,
height=117.8125pt,
width=199.4375pt
]%
{GS5BPB00.wmf}%
\\
Fig. Diagramatic display of the braiding map
\label{fig}%
\end{center}
\end{center}

The smash product and unit, are given respectively by
\begin{align}
\mu_{\substack{\\\Omega^{\prime} }} & \equiv(\mu_{\substack{\\\Omega}%
}\otimes\mu_{\substack{\\\Omega}})\circ(id_{\substack{\\\Omega}}\otimes
\Psi\otimes id_{\substack{\\\Omega}})\text{ ,}\nonumber\\
\text{ }u_{\substack{\\\Omega^{\prime} }}(1)  & =I_{\substack{\\\Omega
^{\prime} }}=I_{\substack{\\\Omega}}\otimes I_{\substack{\\\Omega}}\text{ .
}\label{smab}%
\end{align}
The following lemma \cite{militaru}, which is easily verified, provides with
the compatibility between the above braiding map $\Psi$ and the associative
product $\mu_{\substack{\\\Omega^{^{\prime}} }}$.

\textit{Lemma}: Let $\Omega=\mathcal{A}\otimes\mathcal{B}$ and the linear
braiding map $\Psi:\Omega\otimes\Omega\rightarrow\Omega\otimes\Omega$ given in
eq.(\ref{braidbig}). The algebra $\Omega^{\prime}=\Omega\#_{\substack{\\\Psi
}}\Omega$ is a smash product algebra which implies that the following
relations hold:

i) $\Psi$ is normal i.e%

\begin{align*}
\Psi(x^{^{\alpha}}\otimes\xi^{m}\otimes I_{A}\otimes I_{B})  & =I_{A}\otimes
I_{B}\otimes x^{^{\alpha}}\otimes\xi^{m}\\
\Psi(I_{A}\otimes I_{B}\otimes x^{^{\alpha}}\otimes\xi^{m})  & =(x^{^{\alpha}%
}\otimes\xi^{m}\otimes I_{A}\otimes I_{B})
\end{align*}

ii) $\Psi$ satisfies the following realation:%

\begin{align*}
\Omega^{\otimes4}  & \rightarrow\Omega^{\otimes2}:(I_{\Omega}\otimes
\mu_{\Omega})\circ(\Psi\otimes I_{\Omega})\circ(I_{\Omega}\otimes\mu_{\Omega
}\otimes I_{\Omega})\circ(I_{\Omega}\otimes I_{\Omega}\otimes\Psi)\\
\text{ \ \ }  & =(\mu_{\Omega}\otimes I_{\Omega})\circ(I_{\Omega}\otimes
\Psi)\circ(I_{\Omega}\otimes\mu_{\Omega}\otimes I_{\Omega})\circ(\Psi\otimes
I_{\Omega}\otimes I_{\Omega}).
\end{align*}

The non commutativity of the algebra of increaments can explicitely be seen if
we look at the simplest example of the one-step algebra of increament as follows:

Define the one-step increments $\omega_{ij},\ i,\ j=1,2$ as
\begin{align*}
\omega_{11}  & =x\otimes\xi\otimes I_{A}\otimes I_{B},\ \omega_{12}=x\otimes
I_{B}\otimes I_{A}\otimes\xi,\\
\omega_{21}  & =I_{A}\otimes\xi\otimes x\otimes I_{B},\ \omega_{22}%
=I_{A}\otimes I_{B}\otimes x\otimes\xi,
\end{align*}
then it can be easily checked that the following relations hold:%

\begin{align*}
\omega_{11}\ \omega_{12}  & =qQ\omega_{12}\omega_{11},\ \omega_{21}\omega
_{11}=Q\omega_{11}\omega_{21},\ \omega_{22}\omega_{11}=qQ^{2}\omega_{11}%
\omega_{22},\\
\omega_{12}\omega_{21}  & =q\omega_{21}\omega_{12},\ \omega_{21}\omega
_{22}=qQ\omega_{22}\omega_{21},\ \omega_{22}\omega_{12}=Q\omega_{12}%
\omega_{22}.
\end{align*}

Note that the above choice for $\Psi$ reproduces the braiding of $\mathcal{A}$
and $\mathcal{B}$ algebras. The product $\mu_{\substack{\\\Omega}}$ extends to
an associative product on the n-fold tensor product $\Omega^{n}\equiv
\Omega^{\otimes n}$ , $n\geq2$, by means of the relation%

\begin{equation}
\mu_{\Omega^{n}}=(\mu_{\Omega^{n-1}}\otimes\mu_{\Omega})\circ_{k=1}%
^{2n-3}\circ(id_{\Omega}^{\otimes^{2n-2-k}}\otimes\Psi\otimes id_{\Omega
}^{\otimes k}),\label{miomega}%
\end{equation}
and in this way provides the following braiding relations on $\Omega^{n}:$%
\begin{equation}
x_{i}x_{j}=x_{j}x_{i}\text{, }\forall i,j\text{, }\xi_{i}\xi_{j}=q\xi_{j}%
\xi_{i}\text{, }x_{i}\xi_{j}=Q\xi_{j}x_{i}\text{, for }i>j,\label{rel}%
\end{equation}
where the indices above indicate the position of the embeddings of $x$ and
$\xi$ in the respective spaces (e.g. $\xi_{2}\xi_{1}=(I_{\mathcal{A}}\otimes
I_{\mathcal{B}}\otimes I_{\mathcal{A}}\otimes\xi)(I_{\mathcal{A}}\otimes
\xi\otimes I_{\mathcal{A}}\otimes I_{\mathcal{B}})=q(I_{\mathcal{A}}\otimes
\xi\otimes I_{\mathcal{A}}\otimes I_{\mathcal{B}})(I_{\mathcal{A}}\otimes
I_{\mathcal{B}}\otimes I_{\mathcal{A}}\otimes\xi)=q\xi_{1}\xi_{2}$).

Finally using eqs. (\ref{hopfab},\ref{rel}), we compute the $n$-th fold
coproduct on $x^{k}\otimes{\xi}^{l}$ $\in$ $\Omega$ , $k\in$\textbf{\ }%
$\mathbf{Z}_{+}$ , $l\in\{0,1,...,N-1\}$, which is given by:
\[
\Delta_{\Omega}^{n-1}(x^{k}\otimes{\xi}^{l})=\sum_{i_{1}+\cdots+i_{n}=k}\text{
}\sum_{j_{1}+\cdots+j_{n}=l}\text{ }\left(
\begin{array}
[c]{ccc}%
& k & \\
i_{1} & \cdots &  i_{n}%
\end{array}
\right)  \left[
\begin{array}
[c]{ccc}%
& l & \\
j_{1} & \cdots &  j_{n}%
\end{array}
\right]  _{q}
\]%

\begin{equation}
\times x^{i_{1}}\otimes{\xi}^{j_{1}}\otimes\cdots\otimes x^{i_{n}}\otimes{\xi
}^{j_{n}}\label{ndelta}%
\end{equation}
where $\Delta_{\Omega}^{m}=(\Delta_{\Omega}\otimes id_{\Omega}^{\otimes^{m-1}%
})\circ\Delta_{\Omega}^{m-1}=(id^{\otimes^{m-1}}\otimes\Delta_{\Omega}%
)\circ\Delta_{\Omega}^{m-1}$ , with $\left(
\begin{array}
[c]{ccc}%
& k & \\
i_{1} & \cdots &  i_{n}%
\end{array}
\right)  =\frac{k!}{i_{1}!...i_{n}!}$ , while the $q$-binomial coefficient is
defined as $\left[
\begin{array}
[c]{ccc}%
& l & \\
j_{1} & \cdots &  j_{n}%
\end{array}
\right]  _{q}=\frac{[l]_{q}!}{[j_{1}]_{q}!...[j_{n}]_{q}!}$, where
$[l]_{q}=\frac{q^{l}-1}{q-1}$, $[l]_{q}!=[1]_{q}[2]_{q}...[l]_{q}.$

In the next chapter we will construct a stationary random walk on $\Omega$ and
derive the difussion limit equation of motion.

\section{Diffusion equation on the smash line algebra}

Let $\phi_{\substack{\\\Omega}}:\Omega\longrightarrow\mathbf{C,}$ be a linear
normalized ($\phi_{\substack{\\\Omega}}(\mathbf{1}_{\substack{\\\Omega
}})=1)\;$and positive semi-definite functional from $\Omega$ to $\mathbf{C}$,
which corresponds to probability denisty element $\rho$ satisfying the
following relations:%

\begin{equation}
\phi_{\substack{\\\Omega}}(f)\equiv<f>_{\phi_{\Omega}}=\int\rho f=<\phi
_{\substack{\\\Omega}},f>,\label{phi}%
\end{equation}
where $f=f(x,\xi)$ is a general element (observable) of $\Omega$. It is
assumed that{\ }$\phi_{\substack{\\\Omega}}$ lives in the dual space of
$\Omega$, where the product is defined to be the usual convolution operation
between probability density functions, by means of the following relations:%

\begin{equation}
\phi_{\substack{\\\Omega}}^{\star n}(f)=\text{ }<\phi_{\substack{\\\Omega
}}\star\phi_{\substack{\\\Omega}}\star...\star\phi_{\substack{\\\Omega
}},\text{ }f>\text{ }=\text{ }<\phi_{\substack{\\\Omega}}^{\otimes n},\text{
}\Delta^{n-1}(f)>.\label{nphi}%
\end{equation}
where $\Delta^{n-1}\equiv\Delta_{\Omega}^{n-1}$ will be used thereafter. If
eq.{(\ref{nphi})} is interpreted as the state of probability function of the
random walk after $n$ steps, then we can evaluate the general state after an
n-step walk. Consider a general observable element of $\Omega$ such as%

\begin{equation}
f(x,\xi)=\sum_{l=0}^{N-1}\sum_{k\in\mathbf{Z}_{+}}c_{kl}x^{k}\xi^{l}\equiv
\sum_{l=0}^{N-1}c_{l}(x)\xi^{l}=\sum_{k\in\mathbf{Z}_{+}}x^{k}d_{k}%
(\xi)\label{f}%
\end{equation}
where it should be understood , except if otherwise stated, that $x^{k}%
\equiv(x\otimes I_{B})^{k}=(x^{k}\otimes I_{B})$ and $\xi^{l}\equiv
(I_{A}\otimes{\xi})^{l}=(I_{A}\otimes{\xi}^{l})$. The last equations indicate
that the observable $f(x,\xi)$, can been seen either as an $N-1$-degree
anyonic polynomial with real analytical function as coefficients ($c_{l}(x))$,
or as a real analytic function with anyonically valued coefficients. In view
of the matrix realization of the anyonic variables discussed below, these two
alternatives of representing the general observable quantities of the random
walk might help us to identify a physical model decribed by the smash algebra
$\Omega$, as a model of interacting classical/boson system to a spin like
system. Next using eqs.{(\ref{miomega},\ref{ndelta},\ref{phi},\ref{nphi}%
,\ref{f}), we obtain the convolution
\[
\phi_{\Omega}^{\star n}(f)=
\]
\begin{align}
& \sum_{k,l\in\mathbf{Z}_{+}}\ \sum_{i_{1}+...+i_{n}=k}\ \sum_{j_{1}%
+...+j_{n}=l}\left(
\begin{array}
[c]{ccc}%
& k & \\
i_{1} & \cdots &  i_{n}%
\end{array}
\right)  \left[
\begin{array}
[c]{ccc}%
& l & \\
j_{1} & \cdots &  j_{n}%
\end{array}
\right]  _{q}c{_{kl}}\nonumber\\
& \times{\phi}_{\Omega}(x^{i_{1}}\otimes{\xi}^{j_{1}})...\phi_{\Omega
}(x^{i_{n}}\otimes{\xi}^{j_{n}}).\label{genphi}%
\end{align}
Note that althought $\phi_{\substack{\\\Omega}}$ is linear functional the
convoluted }$\phi_{\substack{\\\Omega}}^{\star n}${\ is a non-linear one. In
order to evaluate the above quantity we can assume that }%

\begin{equation}
{\rho}_{\Omega}{(x,\xi)=\rho}_{A}{(x)\otimes\rho}_{B}{(\xi)}\text{.}%
\label{roinit}%
\end{equation}
{The density matrices $\rho_{A}(x),$ $\rho_{B}(\xi)$ can be taken of the form}%

\begin{align}
\rho_{A}(x)  & =p_{1}\delta(x-a)+(1-p_{1})\delta(x+a),\nonumber\\
\rho_{B}(\xi)  & =p_{2}\delta(\xi-\theta)+(1-p_{2})\delta(\xi+\theta
),\label{chro}%
\end{align}
{where as usual }$p_{1}$ and $p_{2}$ are chosen probabilities{\ and the
anyonic delta function in ${\rho}_{B}(\xi)$ is defined as $\delta(\xi
-\theta)=$}$\sum_{i=0}^{N-1}\theta^{N-1-i}\xi^{i}$ and {$\int{\delta
(\xi+\theta)f(\xi)=f(\theta)}$\cite{majidplaza}}. {This choice allows us to
determine explicitly average values of the generating monomials of the algebra
after $n $ steps. Note that the value of the size of each random step in
}$\Omega$ is considered in general to be $a\otimes\theta${\ . For the random
walk on the real line in }$\Omega$ we have that{\
\begin{align}
& <x^{m}>_{{\phi}_{\Omega}^{\star n}}=\sum_{s_{1}+...+s_{n}=m}\left(
\begin{array}
[c]{ccc}%
& m & \\
s_{1} & \cdots &  s_{n}%
\end{array}
\right)  \Pi_{l=1}^{n}<x^{s_{l}}>_{\substack{\\\phi_{A} }}\nonumber\\
& =\sum_{s_{1}+...+s_{n}=m}\left(
\begin{array}
[c]{ccc}%
& m & \\
s_{1} & \cdots &  s_{n}%
\end{array}
\right)  \Pi_{l=1}^{n}(p_{1}a^{s_{l}}+(1-p_{1})(-a)^{s_{l}})\label{phix}\\
& =\sum_{s_{1}+...+s_{n}=m}\left(
\begin{array}
[c]{ccc}%
& m & \\
s_{1} & \cdots &  s_{n}%
\end{array}
\right)  \Pi_{l=1}^{n}\left[  p_{1}e^{aD_{x}}\otimes\varepsilon_{\mathcal{B}%
}+(1-p_{1})e^{-aD_{x}}\otimes\varepsilon_{\mathcal{B}}\right]  _{\mid_{x=0}%
}(x^{s_{l}}\otimes I_{\mathcal{B}})\nonumber\\
& =\sum_{s_{1}+...+s_{n}=m}\left(
\begin{array}
[c]{ccc}%
& m & \\
s_{1} & \cdots &  s_{n}%
\end{array}
\right)  \Pi_{l=1}^{n}\phi_{A}(x^{s_{l}})\nonumber\\
& =\sum_{s_{1}+...+s_{n}=m}\left(
\begin{array}
[c]{ccc}%
& m & \\
s_{1} & \cdots &  s_{n}%
\end{array}
\right)  \Pi_{l=1}^{n}<x^{s_{l}}>_{\phi_{A}}\nonumber
\end{align}
where $D_{x}=\partial/\partial_{x}$. }

With respect to random walks on the anyonic line in $\Omega$ we have that:
\begin{align}
& <\xi^{t}>_{\substack{\\{\phi}_{\Omega}^{\star n} }}=\sum_{r_{1}+...+r_{n}%
=t}\left[
\begin{array}
[c]{ccc}%
& t & \\
r_{1} & \cdots &  r_{n}%
\end{array}
\right]  _{q}\Pi_{l=1}^{n}<\xi^{r_{l}}>_{\substack{\\\phi_{B} }}\nonumber\\
& =\sum_{r_{1}+...+r_{n}=t}\left[
\begin{array}
[c]{ccc}%
& t & \\
r_{1} & \cdots &  r_{n}%
\end{array}
\right]  _{q}\Pi_{l=1}^{n}\left(  p_{2}\theta^{r_{l}}+(1-p_{2})(-\theta
)^{r_{l}}\right) \nonumber\\
& =\text{ }\sum_{r_{1}+...+r_{n}=t}\left[
\begin{array}
[c]{ccc}%
& t & \\
r_{l} & \cdots &  r_{n}%
\end{array}
\right]  _{q}\Pi_{l=1}^{n}\left[  \varepsilon_{\mathcal{A}}\otimes
p_{2}e^{\theta D_{\xi}}+(1-p_{2})\varepsilon_{\mathcal{A}}\otimes
e^{-\theta{D_{\xi}}}\right]  _{\mid_{\xi=0}}(I_{\mathcal{A}}\otimes\xi^{r_{l}%
})\nonumber\\
& =\text{ }\sum_{r_{1}+...+r_{n}=t}\left[
\begin{array}
[c]{ccc}%
& t & \\
r_{l} & \cdots &  r_{n}%
\end{array}
\right]  _{q}\Pi_{l=1}^{n}\phi_{B}(\xi^{r_{l}})\label{phixi}\\
& =\text{ }\sum_{r_{1}+...+r_{n}=t}\left[
\begin{array}
[c]{ccc}%
& t & \\
r_{l} & \cdots &  r_{n}%
\end{array}
\right]  _{q}\Pi_{l=1}^{n}<\xi^{r_{l}}>_{\substack{\\\phi_{B} }}\nonumber
\end{align}
where the derivative $D_{\substack{\\\xi}}$ is defined {in
\cite{majidplaza,exton} }as {$D_{\substack{\\\xi}}f(\xi)=\frac{f(\xi)-f(\xi
q)}{(1-q)\xi}$ , and satisfies the }$q${-Leibniz rule $D_{\substack{\\\xi
}}(fg)=(D_{\substack{\\\xi}}f)g+($}$L_{q}f)D_{\substack{\\\xi}}g$, where
$L_{q}f(\xi)=f(q\xi)${. }

Finally with respect to simultaneous random walks in both $x,$ $\xi$
directions we obtain using eqs.(\ref{phix},\ref{phixi}):%

\begin{align}
& <x^{k}\otimes{\xi}^{l}>_{{\phi}^{\star n}}={\phi}^{\star n}(x^{k}\otimes
{\xi}^{l})\nonumber\\
& =\sum_{i_{1}+..+i_{n}=k}{\text{ }\sum_{j_{1}+...+j_{n}=l}\text{ }}\left(
\begin{array}
[c]{ccc}%
& k & \\
i_{1} & ... & i_{n}%
\end{array}
\right)  \left[
\begin{array}
[c]{ccc}%
& l & \\
j_{1} & ... & j_{n}%
\end{array}
\right]  _{q}\label{pfifull}\\
& \times\prod_{s=1}^{n}\phi_{x}(x{^{i_{s}}})\phi_{\substack{\\\xi}%
}({\xi^{j_{s}}})\\
& =\phi_{x}^{\star n}(x^{k})\phi_{\substack{\\\xi}}^{\star n}({\xi}%
^{l})\nonumber
\end{align}
where $\phi_{x}(x{^{i_{s}}})\phi_{\substack{\\\xi}}({\xi^{j_{s}}})=<x{^{i_{s}%
}}>_{\phi_{x}}<{\xi^{j_{s}}}>_{\phi_{\substack{\\\xi}}}$.

{Let us now compute the system after }$n${\ steps and its limit }as
$n\rightarrow\infty$ . {Using Taylor's expansion, for the form of }%
$\phi_{\substack{\\\Omega}}^{\star n}$ as it can be easily read off from
equations {(\ref{nphi},\ref{phix},\ref{phixi}), and (\ref{pfifull}) we have}
{\ that
\begin{align}
{\phi}_{\Omega}^{\star n}  & ={\phi}_{A}^{\star n}\otimes{\phi}_{B}^{\star
n}=([\varepsilon_{\mathcal{A}}+2a(p_{1}-1/2)D_{x}+a^{2}/2!D_{x}^{2}%
+...]^{n}\label{phin}\\
& \otimes[\varepsilon_{\mathcal{B}}+2\theta(p_{2}-1/2)D_{\substack{\\\xi
}}+{\theta}^{2}/[2]_{q}!D_{\substack{\\\xi}}^{2}+...]^{n})\mid_{x=0,\xi
=0}.\nonumber
\end{align}
Following \cite{majid,majidplaza}, we make the following substitutions}%

\begin{align}
2a(p_{1}-1/2)  & =\frac{c_{1}t}n,\text{ }2\theta(p_{2}-1/2)=\frac{c_{2}%
t}n,\nonumber\\
a^{2}/2  & =\frac{{\alpha}_{1}t}n,\text{ }\theta^{2}/[2]_{q}=\frac{{\alpha
}_{2}t}n.\label{substit}%
\end{align}
{\ Then we take the limit $n\rightarrow\infty$ with $t$, $c_{1}$, $c_{2}$,
${\alpha}_{1}$, ${\alpha}_{2}$ kept fixed and $t=nd$, }$d${\ being the size of
the step in time (which in the limit taken }$d${$\rightarrow0$ as }$a$ does{),
to obtain the continue limit of random walk where the steps are viewed as
steps in time. Thus we obtain: \
\begin{equation}
{\phi}_{\Omega}^{\infty}(f)=(e^{(c_{1}tD_{x}+{\alpha}_{1}tD_{x}^{2}%
)\otimes\varepsilon_{\mathcal{B}}+\varepsilon_{\mathcal{A}}\otimes
(c_{2}tD_{\substack{\\\xi}}+{\alpha}_{2}tD_{\substack{\\\xi}}^{2})}%
f)_{\mid_{x,\xi=0},}\label{phiapir}%
\end{equation}
where the limit $(1+z/n)^{n}\rightarrow e^{z}$ for $n\rightarrow\infty$, has
been used. The associated density $\rho_{\Omega}^{\infty}$ can now be obtained
by evaluating }%

\[
{{\phi}_{\Omega}^{\infty}(\delta)=\int\rho^{\infty}\delta(x,\xi)}\text{, where
}\delta{(x,\xi)\equiv\delta(x-a)\otimes\delta(\xi-\theta).}
\]
{Then\
\begin{equation}
\rho_{\Omega}^{\infty}(a,\theta)=\rho_{A}^{\infty}(a)\otimes\rho_{B}^{\infty
}(\theta),\label{ro}%
\end{equation}
}where
\begin{align}
\rho_{A}^{\infty}(a)  & =(4{\pi}{\alpha}_{1}t)^{-1}e^{-\frac{(a-c_{1}t)^{2}%
}{4{\alpha}_{1}t}},\label{rox}\\
\rho_{B}^{\infty}(\theta)  & =\sum_{k=0}^{N-1}{\theta}^{N-1-k}\sum
_{l=0}^{l<k/2}\frac{(c_{2}t)^{k-l}({\alpha}_{2}/c_{2})^{l}[k]_{q}!}%
{l!(k-2l)!}.\label{roxi}%
\end{align}
To obtain the diffusion equation we take a generic $f$ of the form {(\ref{f})
and evaluate: }

{\
\begin{align}
\int(\partial_{t}\rho_{\Omega}^{\infty})f  & =\partial_{t}\phi_{\Omega
}^{\infty}(f)=\phi_{\Omega}^{\infty}((c_{1}D_{x}+{\alpha}_{1}D_{x}^{2}%
+c_{2}D_{\substack{\\\xi}}+{\alpha}_{2}D_{\substack{\\\xi}}^{2})f)\nonumber\\
& =\int\rho_{\Omega}^{\infty}(c_{1}D_{x}+{\alpha}_{1}D_{x}^{2}+c_{2}%
D_{\substack{\\\xi}}+{\alpha}_{2}D_{\substack{\\\xi}}^{2})f\label{prodif}\\
& =\int((-c_{1}D_{x}+{\alpha}_{1}D_{x}^{2}+c_{2}D_{\substack{\\\xi}%
}^{*}+{\alpha}_{2}D _{\substack{\\\xi}}^{*2})\rho^{\infty})f,\nonumber
\end{align}
where we have used the definition of }$D_{\substack{\\\xi}}^{*}%
=-D_{\substack{\\\xi}}L_{q^{-1}}$, $L_{q^{-1}}$ being such that $L_{q^{-1}%
}f(\xi)=f(q^{-1}\xi)$. Then eq.(\ref{prodif}) {leads to the desired diffusion
equation:
\begin{equation}
\partial_{t}\rho_{\Omega}^{\infty}=((-c_{1}D_{x}+{\alpha}_{1}D_{x}^{2})\otimes
id_{\mathcal{B}}+id_{\mathcal{A}}\otimes(c_{2}D_{\substack{\\\xi}}^{*}%
+{\alpha}_{2}D_{\substack{\\\xi}}^{*2}))\rho_{\Omega}^{\infty}\label{dif1}%
\end{equation}
}whose solution is given in eqs.(\ref{ro},\ref{rox},\ref{roxi}) .

Before closing this chapter we will quote the generating function of the
statistical moments of the two random variables, which are:%

\[
G(k_{1},k_{2})\equiv<e^{ik_{1}x}\otimes e_{q}^{ik_{2}\xi}>_{\substack{\\\phi
}}=\sum_{m_{1\geq0}}\sum_{m_{2}=0}^{N-1}\frac{(ik_{1})^{m_{1}}}{m_{1}!}%
\frac{(ik_{2})^{m_{2}}}{m_{2}!}<x^{m_{1}}>_{\substack{\\\phi}}<\xi^{m_{2}%
}>_{\substack{\\\phi}}
\]
Then we can obtain the moments as:
\begin{align}
\frac{d^{l}}{i^{l}dk_{1}^{l}}G(k_{1},k_{2})_{\mid_{k_{1}=k_{2}=0}}  &
=<x^{l}>_{\substack{\\\phi}},\nonumber\\
(1/i)D_{q,k_{2}}G(k_{1},k_{2})_{\mid k_{1}=k_{2}=0}  & =<\xi
>_{\substack{\\\phi}},\nonumber\\
(1/i^{l})D_{q,k_{2}}^{l}G(k_{1},k_{2})_{\mid_{k_{1}=k_{2}=0}}  & =<\xi
^{l}>_{\substack{\\\phi}},
\end{align}
where $D_{q,k_{2}}f(k_{2})=\frac{f(k_{2})-f(qk_{2})}{(1-q)k_{2}}$, the
\textit{q}-deformed derivative\cite{exton}.

\section{\textbf{Non stationary random walk}}

Consider now the Hamiltonian evolution of a quantity $F=F(x,p,\xi
,p_{\substack{\\\xi}}),$ depending in general on the phase space variables
$x$, $p$ and $\xi$, $p_{\substack{\\\xi}}$ dermined by the following Poisson brackets%

\begin{align}
\{F,H\}  & =\frac{\partial F}{\partial x}\frac{\partial H}{\partial p}%
-\frac{\partial F}{\partial p}\frac{\partial H}{\partial x}+(-1)^{^{\epsilon}%
}(\frac{\partial F}{\partial\xi}\frac{\partial H}{\partial p_{\xi}}%
+\frac{\partial F}{\partial p_{\xi}}\frac{\partial H}{\partial\xi}%
)=V_{H}(F)\nonumber\\
& =V_{H}^{1}(F)+V_{H}^{2}(F),\label{vh}%
\end{align}
where we have assumed that $\xi^{2}=0.$ The parity degree of $F,\deg
(F)=\epsilon$ takes values $\epsilon=0,1$ to account of the even or odd
chracter of $F$ respectively. The form and action of the operators $V_{H}%
^{1}=\frac{\partial H}{\partial p}\frac\partial{\partial x}-\frac{\partial
H}{\partial x}\frac\partial{\partial p}$, $V_{H}^{2}$ =$(-1)^{^{\epsilon}%
}(\frac\partial{\partial\xi}\frac{\partial H}{\partial p_{\xi}}+\frac
\partial{\partial p_{\xi}}\frac{\partial H}{\partial\xi}),\ $can be read off
from the above equation. We can thus defined, using $t$ as a time parameter,
the action of an evolution operator $e^{tV_{H}}$ on $x_{0}$ , by $e^{tV_{H}%
}x_{0}=x_{t}$, and on $\xi_{0}$ by $e^{tV_{H}}\xi_{0}=\xi_{t}$ where $x_{t}$
and $\xi_{t}$ are the values of the observables at time $t$ . We shall now
proceed to calculate moments after one step walk in $\Omega$, $\phi(x_{t}%
^{k}\otimes\xi_{t}^{l}),$ where $k\in\mathbf{Z}_{+}$ , $l=0,1$, using the
probability densities for time $t=0$, given as $\rho_{0}^{1}(x_{0}%
)=p_{1}\delta(x_{0}-a)+(1-p_{1})\delta(x_{0}+a)$, $\rho_{0}^{2}(\xi_{0}%
)=p_{2}\delta(\xi_{0}-\theta)+(1-p_{2})\delta(\xi_{0}+\theta)$, while the
delta functions for the $\xi$ variables are as defined previously. Moreover we
shall again assume that $\rho_{0}(x_{0},\xi_{0})=\rho_{0}^{1}(x_{0}%
)\otimes\rho_{0}^{2}(\xi_{0}).$ The reason for these choices will become
apparent bellow. Thus,%

\begin{align}
\phi(x_{t}\otimes\xi_{t})  & =<\rho_{0},e^{tV_{H}^{1}}x_{0}\otimes
e^{tV_{H}^{2}}\xi_{0}>\nonumber\\
& =<e^{-tV_{H}^{1}}\rho_{0}^{1},x_{0}><e^{-tV_{H}^{2}}\rho_{0}^{2},\xi
_{0}>=\rho^{1}(x_{t})\rho^{2}(\xi_{t})\nonumber\\
& =\rho_{t}^{1}(x_{0})\rho_{t}^{2}(\xi_{0}).\label{phiti}%
\end{align}
Direct evaluation of $\rho_{t}^{1}$ and $\rho_{t}^{2}$ gives%

\begin{align}
\rho_{t}^{1}(x_{0})  & =p_{1}\delta(x_{0}-t\frac{\partial H}{\partial
p}-a)+(1-p_{1})\delta(x_{0}-t\frac{\partial H}{\partial p}+a)\nonumber\\
& =p_{1}\delta(x_{0}-t\lambda-a)+(1-p_{1})\delta(x_{0}-t\lambda
+a),\label{roxt}\\
\rho_{t}^{2}(\xi_{0})  & =p_{2}\delta(\xi_{0}-t\frac{\partial H}{\partial
p_{\substack{\\\xi}}}-\theta)+(1-p_{2})\delta(\xi_{0}-t\frac{\partial
H}{\partial p_{\substack{\\\xi}}}+\theta)\\
& =p_{2}\delta(\xi_{0}-t\widetilde{\lambda}-\theta)+(1-p_{2})\delta(\xi
_{0}-t\widetilde{\lambda}+\theta).
\end{align}
where $\lambda=\frac{\partial H}{\partial p}$ and $\tilde{\lambda}%
=-\frac{\partial H}{\partial p_{\substack{\\\xi}}}$ and thus, using
eqs.(\ref{phi},\ref{phiti}) we finally have%

\begin{align}
\phi(x_{t}\otimes\xi_{t})  & =[(p_{1}e^{a_{t}D_{x_{0}}}\otimes\varepsilon
_{\mathcal{B}}+(1-p_{1})e^{a_{t}^{\prime}D_{x_{0}}}\otimes\varepsilon
_{\mathcal{B}})\nonumber\\
& \times(\varepsilon_{\mathcal{A}}\otimes p_{2}e^{\theta_{t}D_{\xi_{0}}%
}+(1-p_{2})\varepsilon_{\mathcal{A}}\otimes e^{\theta_{t}^{\prime}D_{\xi_{0}}%
})]_{\mid_{x_{0},\xi_{0}=0}}(x_{0}\otimes\xi_{0})\nonumber\\
& =e^{-\lambda t\partial/\partial a}e^{-\widetilde{\lambda}t\partial
/\partial\theta}\phi_{t=0}(x_{0}\otimes\xi_{0})\label{phitout}\\
& =\phi_{t}(x_{0}\otimes\xi_{0})\nonumber
\end{align}
where we have set $a_{t}=a-t\lambda$ , $a_{t}^{\prime}=-a-t\lambda$,
$\theta_{t}=\theta-t\tilde{\lambda}$ and $\theta_{t}^{\prime}=-\theta
-t\tilde{\lambda}$ . Evaluation of $\phi(f)$ for a generic $f$ of the form
given above can be straightforwardly calculated using eq.(\ref{phitout}). To
derive the continue limit we proceed as before. By Taylor's expansion of
$\phi_{t}$ and taking ${\phi}_{t}^{\star n}$ we have
\begin{align}
\phi_{t}^{\star n}  & =[\varepsilon_{\mathcal{\Omega}}+(2a(p_{1}%
-1/2)-t\lambda)D_{x}\otimes\varepsilon_{\mathcal{B}}+\nonumber\\
& (\frac{a^{2}}2+\frac{(t\lambda)^{2}}2+at\lambda-2p_{1}ta\lambda)D_{x}%
^{2}\otimes\varepsilon_{\mathcal{B}}+...]^{n}\nonumber\\
& \times[\varepsilon_{\mathcal{\Omega}}+\varepsilon_{\mathcal{A}}%
\otimes(2\theta(p_{2}-1/2)-t\widetilde{\lambda})D_{\substack{\\\xi
}}+\nonumber\\
& \varepsilon_{\mathcal{A}}\otimes(\frac{\theta^{2}}2+\frac{(t\widetilde
{\lambda})^{2}}2+\theta t\widetilde{\lambda}-2p_{2}t\theta\widetilde{\lambda
})D_{\substack{\\\xi}}{^{2}}+...]^{n}.\label{phint}%
\end{align}
As before we are making the following substitutuions%

\begin{align}
2a(p_{1}-1/2)  & =\frac{c_{1}t}n\text{, }t\lambda=\lambda\frac{d_{1}t}n\text{,
}\frac{a^{2}}2=\frac{\alpha_{1}t}n,\nonumber\\
2\theta(p_{2}-1/2)  & =\frac{c_{2}t}n\text{, }t\tilde{\lambda}=\tilde{\lambda
}\frac{d_{2}t}n\text{, }\frac{\theta^{2}}2=\frac{\alpha_{2}t}n,\label{subs2}%
\end{align}
where $d_{1}$, $d_{2}$, real constants. Here $D_{\substack{\\\xi}}^{2}=0$ and
thus the last of eq.(\ref{subs2}) is not needed. Taking the limit
$n\rightarrow\infty$ as before we obtain for a generic function $f(x,\xi)$%

\begin{align}
{\phi}^{\infty}(f)  & =[e^{((c_{1}-\lambda d_{1})tD_{x}+{\alpha}_{1}tD_{x}%
^{2})\otimes\varepsilon_{\mathcal{B}}+\varepsilon_{\mathcal{A}}\otimes
(c_{2}-\tilde{\lambda}d_{2})tD_{\substack{\\\xi}}}f)_{_{\mid x,\xi=0}%
}\nonumber\\
& =(e^{tK}f)_{_{\mid x,\xi=0}}=\int\rho^{\infty}f=<\rho^{\infty}%
,f>.\label{phitinf}%
\end{align}
The diffusion equation is immediately implied from the fact that:
\[
<\partial_{t}\rho^{\infty},f>=\int\partial_{t}\rho^{\infty}f=(e^{tK}Kf)_{\mid
x,\xi=0}=<\rho^{\infty},Kf>=<K^{*}\rho^{\infty},f>
\]
giving%

\begin{align}
\frac{\partial\rho^{\infty}}{\partial t}  & =[-c_{1}D_{x}\otimes
id_{\mathcal{B}}+{\alpha}_{1}D_{x}^{2}\otimes id_{\mathcal{B}}+id_{\mathcal{A}%
}\otimes c_{2}D_{\substack{\\\xi}}^{*}\nonumber\\
& +\frac{\partial H}{\partial p}D_{x}\otimes id_{\mathcal{B}}-id_{\mathcal{A}%
}\otimes(-\frac{\partial H}{\partial p_{\substack{\\\xi}}})D_{\substack{\\\xi
}}^{*}]\rho^{\infty}\label{difg}%
\end{align}
where $D_{\substack{\\\xi}}^{*}$ is defined by $D_{\substack{\\\xi}}^{*}=-D
_{\substack{\\\xi}}L_{q^{-1}}$ ( $L_{q^{-1}}f(\xi)=f(q^{-1}\xi)$ for any
anyonic function $f(\xi)$ ) and where we have set $d_{1}=d_{2}=1$. It should
be pointed out though that we can express eqs.(\ref{dif1}, \ref{difg}) using
the finite dimensional representation of $\xi$, $D_{\substack{\\\xi}}$ as we
shall see in the next section.

\section{Matrix realization of random walk and diffusion}

We will now implement the $N$ -dimentional matrix representations of $\xi$,
$D_{\substack{\\\xi}}$, $D_{\substack{\\\xi}}^{*}$ as they have explicitely
been constructed in \cite{rausch} . As to matrix representation of
$D_{\substack{\\\xi}}^{*}$ one should take in to account the so called anyonic
Leibnitz rule $D_{\substack{\\\xi}}(fg)=(D_{\substack{\\\xi}}f)g+L_{q}%
f(D_{\substack{\\\xi}}g)$ and the property that $D_{\substack{\\\xi}%
}^{*}=-D_{\substack{\\\xi}}L_{q^{-1}}$\cite{majidplaza,elltso}. Thus we have that,%

\[
\xi=\left(
\begin{array}
[c]{ccccccc}%
0 & 0 & 0 & 0 & 0 & 0 & ...\\
1 & 0 & 0 & 0 & 0 & 0 & ...\\
0 & 1 & 0 & . & . & . & ...\\
. & . & . & . & . & . & .\\
. & . & . & . & . & 1 & 0
\end{array}
\right)  ,\text{ }D_{\substack{\\\xi}}=\left(
\begin{array}
[c]{ccccccc}%
0 & \{1\} & 0 & 0 & 0 & 0 & ...\\
. & 0 & \{2\} & 0 & 0 & 0 & ...\\
. & . & . & . & . & . & ...\\
. & . & . & . & . & 0 & \{N-1\}\\
. & . & . & . & . & . & 0
\end{array}
\right)  ,
\]%

\begin{equation}
D_{\substack{\\\xi}}^{*}=-\left(
\begin{array}
[c]{ccccccc}%
0 & \{1\}e^{\omega^{-1}\{1\}} & 0 & 0 & 0 & 0 & ...\\
. & 0 & \{2\}e^{\omega^{-1}\{2\}} & 0 & 0 & 0 & ...\\
. & . & . & . & . & . & ...\\
. & . & . & . & . & 0 & \{N-1\}e^{\omega^{-1}\{N-1\}}\\
. & . & . & . & . & . & 0
\end{array}
\right)  ,\label{dxi}%
\end{equation}
where $\{x\}=\frac{1-\omega^{x}}{1-\omega}$, $\omega=e^{2i\pi/N}$ . The matrix
representation of higher powers of $D_{\substack{\\\xi}}$, $D_{\substack{\\\xi
}}^{*}$ and $\xi$ can easily been obtained. For a general $\rho(x,\xi)$ we
have the lower triangular form%

\begin{align}
\rho(x,\xi,t)  & =\sum_{i=0}^{N-1}\sum_{j=0}^{\infty}\rho_{ij}(t)x^{j}%
\otimes\xi^{i}=\sum_{i=0}^{N-1}\rho_{i}(x,t)\otimes\xi^{i}\nonumber\\
& =\left(
\begin{array}
[c]{ccccccc}%
\rho_{0}(x,t) & 0 & 0 & 0 & 0 & 0 & ...\\
\rho_{1}(x,t) & \rho_{0}(x,t) & 0 & 0 & 0 & 0 & ...\\
\rho_{2}(x,t) & \rho_{1}(x,t) & \rho_{0}(x,t) & . & . & . & ...\\
. & . & . & . & . & . & .\\
\rho_{N-1}(x,t) & \rho_{N-2}(x,t) & . & . & . & \rho_{1}(x,t) & \rho_{0}(x,t)
\end{array}
\right)  .\label{rogen}%
\end{align}
For stationary random walks and general $N$-potent variables, we can now
express eq.(\ref{dif1}) in matrix form, using eqs.(\ref{dxi},\ref{rogen}). The
right hand side of eq.(\ref{dif1}) becomes an operator valued upper band
matrix :%

\begin{equation}
\left[
\begin{array}
[c]{cccccccc}%
H_{x} & c_{2}\lambda_{1} & \alpha_{2}\lambda_{1}\lambda_{2} & 0 & 0 & 0 & 0 &
0\\
0 & H_{x} & c_{2}\lambda_{2} & \alpha_{2}\lambda_{2}\lambda_{3} & 0 & 0 & 0 &
0\\
0 & 0 & H_{x} & c_{2}\lambda_{3} & \alpha_{2}\lambda_{3}\lambda_{4} & 0 & 0 &
0\\
0 & 0 & 0 & H_{x} & c_{2}\lambda_{4} & \ddots & \vdots & \vdots\\
0 & 0 & 0 & 0 & H_{x} & \ddots & \ddots & 0\\
\vdots & \vdots & \vdots & \vdots & \vdots & \ddots &  c_{2}\lambda_{N-2} &
\alpha_{2}\lambda_{N-2}\lambda_{N-1}\\
0 & 0 & 0 & 0 & 0 & 0 & H_{x} & c_{2}\lambda_{N-1}\\
0 & 0 & 0 & 0 & 0 & 0 & 0 & H_{x}%
\end{array}
\right]  ,\label{h}%
\end{equation}
where $\lambda_{i}=\{i\}e^{\omega^{-1}\{i\}}$ , $i=1,\cdots,N-1$,
$H_{x}=-c_{1}D_{x}+{\alpha}_{1}D_{x}^{2}$ , while in the left hand side
$\rho^{\infty}(x,\xi,t)$ is as in eq.(\ref{rogen}). The system of differential
equations obtained from eqs.(\ref{rogen},\ref{h}), discribing the continue
limit of stationary random walk in $\Omega,$ can be derived and is given by%

\begin{equation}
\frac{\partial\rho_{k}}{\partial t}=H_{x}\rho_{k}+c_{2}\lambda_{k+1}\rho
_{k+1}+\alpha_{2}\lambda_{k+1}\lambda_{k+2}\rho_{k+2}\label{statsyst}%
\end{equation}
for\ \ \ k=0,1,...,N-1\ .The generalization to the non stationary case is a
straightforward application of the above matrix representations and the
relations in eq.(\ref{difg}) and will not be dealt with explicitly here.
Rather we shall give an explicit solution of eq.(\ref{statsyst}) for the case
$N=2,$ where the above equation takes the form%

\begin{equation}
\frac\partial{\partial t}\left[
\begin{array}
[c]{l}%
\rho_{0}(x,t)\\
\rho_{1}(x,t)
\end{array}
\right]  =\left[
\begin{array}
[c]{ll}%
H_{x} & -c_{2}\\
0 & H_{x}%
\end{array}
\right]  \left[
\begin{array}
[c]{l}%
\rho_{0}\\
\rho_{1}%
\end{array}
\right]  ,\label{n2ro}%
\end{equation}
which after integration gives the time evolution%

\begin{equation}
\left[
\begin{array}
[c]{l}%
\rho_{0}(x,t)\\
\rho_{1}(x,t)
\end{array}
\right]  =(e^{-tH_{x}}+\left[
\begin{array}
[c]{ll}%
1 & -tc_{2}\\
0 & 1
\end{array}
\right]  )\left[
\begin{array}
[c]{l}%
\rho_{0}(x,0)\\
\rho_{1}(x,0)
\end{array}
\right]  .\label{intro}%
\end{equation}
If initially $\rho_{i}(x,0)=f_{i}(x)=\sum_{n\geq0}f_{i,n}x^{n}$ , $i=0,1$,
then we obtain that%

\begin{equation}
\rho_{i}^{^{\prime}}(x,t)=e^{-tH_{x}}\rho_{i}(x,0)=\sum_{n\geq0}f_{i,n}%
H_{n}(x-tc_{1},t\alpha_{1}),\label{sol}%
\end{equation}
where the two variable generalized Hermite polynomial $H_{n}(x,y)=e^{y\frac
{\partial^{2}}{\partial x^{2}}}(x^{n})$ has been used. These polynomials have
the generating function $e^{xt+yt^{2}}=\sum_{n\geq0}\frac{t^{n}}{n!}%
H_{n}(x,y)$, and the expansion $H_{n}(x,y)=n!\sum_{r=0}^{[n/2]}\frac
{y^{r}x^{n-2r}}{r!(n-2r)!}$. Interestingly enough these polynomials can been
generated by a pair of step operators $a^{+}$, $a^{-}$ as%

\[
a^{+}H_{n}(x,y)=H_{n+1}(x,y),\ a^{-}H_{n}(x,y)=nH_{n-1}(x,y).
\]
where $a^{+}$, $a^{-}$ satisfy the Heisenberg canonical commutation relation
$[a^{-},a^{+}]=1$ and admit the realization $a^{+}=x+2xy\frac{\partial^{2}%
}{\partial x^{2}}$, $a^{-}=\frac\partial{\partial x}\;$(see e.g .
\cite{dattoli}).

Finally the solution of eq.(\ref{n2ro}) reads%

\[
\left[
\begin{array}
[c]{l}%
\rho_{0}(x,t)\\
\rho_{1}(x,t)
\end{array}
\right]  =\left[
\begin{array}
[c]{l}%
\rho_{0}^{^{\prime}}(x,t)+\rho_{0}(x,0)-tc_{2}\rho_{1}(x,0)\\
\rho_{1}^{^{\prime}}(x,t)+\rho_{1}(x,0)
\end{array}
\right]  .
\]

\section{Transition operators}

In this chapter we shall formulate Markov processes on the smash line algebra.
The key point is to define the Markov transition operator $T_{\substack{\\\phi
}}:\Omega\rightarrow\Omega,$ where $\phi$ is as before a linear functional on
$\Omega$, such that%

\begin{equation}
T_{\substack{\\\phi}}=(\phi\otimes id_{\Omega}){\Delta}_{{\Omega}%
}{,\ \ \ \varepsilon}_{\substack{\\{\Omega} }}{\circ} T_{\substack{\\\phi
}}=\phi,\ \ \ \phi*\psi={\varepsilon}_{\substack{\\\Omega}}{\circ}%
T_{\substack{\\\psi}}T_{\substack{\\\phi}}.\label{markt}%
\end{equation}
Last relation implies that the Markov transition operators
$T_{\substack{\\\phi}},$ can be composed to form a semigroup with
$T_{\phi=\varepsilon_{\Omega}}=id_{\substack{\\\Omega}},$ as unit element,
which is homomorphic to the convolution semigroup of functionals with unit
element the counit map ${\varepsilon}_{\substack{\\\Omega}},$ and generator
the functional $\phi.\;$In this active picture of random walk in terms of
transition operators, to find the expectation values of the observables after
$n$ steps we employ the relation
\[
\langle f\rangle_{\phi_{1}\star\phi_{2}\star\ldots\star\phi_{n}}%
=\varepsilon_{\substack{\\\Omega}}\circ T_{\phi_{1}}T_{\phi_{2}}\ldots
T_{\phi_{n}}(f),
\]
where in general the states $\phi$ are varying at each step. Let $f(x,\xi)$ be
any function smooth of $x$ and $\xi$ . We shall look at stationary random
walks first. Using eq.(\ref{markt}) with $\phi(f)=\int\rho f$ , and
probability density as $\rho(x,\xi)=\rho_{1}(x)\otimes\rho_{2}(\xi),$ with
$\rho_{1}(x),$ $\rho_{2}(\xi)$ as in eq.(\ref{chro}), we obtain that after one step%

\begin{align}
(T_{\substack{\\\phi}}f)(x,\xi)  & ={p_{1}p_{2}}f({x+a,\xi+\theta}%
)+{(1-p_{1})p_{2}}f({x-a,\xi+\theta})\nonumber\\
& +{p_{1}}(1-{p_{2}})f({x+a,\xi-\theta})+(1-{p_{1}})(1-{p_{2}})f({x-a,\xi
-\theta})\nonumber\\
& =\left[  {p_{1}e}^{aD_{x}}\otimes id_{\mathcal{B}}+(1-{p_{1}}){e}^{-aD_{x}%
}\otimes id_{\mathcal{B}}\right] \nonumber\\
& \times\left[  {p_{2}}id_{\mathcal{A}}\otimes{e}^{\theta D_{\substack{\\\xi
}}}+(1-{p_{2}})id_{\mathcal{A}}\otimes{e}^{-\theta D_{\substack{\\\xi}%
}}\right]  f(x,\xi)\label{t1}\\
& =(T_{\phi_{x}}\otimes T_{\substack{\phi_{\xi}\\}})f(x,\xi),
\end{align}
while when $\rho(x,\xi)=\frac12[\rho_{1}(x)\otimes I_{\mathcal{B}%
}+I_{\mathcal{A}}\otimes\rho_{2}(\xi)],$ we obtain%

\begin{align}
(T_{\substack{\\\phi}}f)(x,\xi)  & ={p_{1}}f({x+a,\xi})+{(1-p_{1})}f({x-a,\xi
})+\nonumber\\
& {p_{2}f({x,\xi+\theta})+(1-p_{2})}f({x,\xi-\theta})\nonumber\\
& =\left[  {p_{1}e}^{aD_{x}}\otimes id_{\mathcal{B}}+(1-{p_{1}}){e}^{-aD_{x}%
}\otimes id_{\mathcal{B}}\right] \label{t2}\\
& +\left[  {p_{2}}id_{\mathcal{A}}\otimes{e}^{\theta D_{\substack{\\\xi}%
}}+(1-{p_{2}})id_{\mathcal{A}}\otimes{e}^{-\theta D_{\substack{\\\xi}%
}}\right]  f(x,\xi)\nonumber\\
& =(T_{\phi_{x}}\otimes id_{\mathcal{B}}+id_{\mathcal{A}}\otimes T_{\phi_{\xi
}})f(x,\xi).
\end{align}
The Markov transition operator in the continue limit $T_{\substack{\\\phi
^{\infty}}}$ can be obtained from eq.(\ref{t1}) with exactly the same
procedure of Taylor expansion and the use of substitutions as in
eq.(\ref{substit}), this yields the operator%

\begin{equation}
T_{\phi^{\infty}}=e^{(-c_{1}tD_{x}+{\alpha}_{1}tD_{x}^{2})\otimes
id_{\mathcal{B}}+id_{\mathcal{A}}\otimes(c_{2}tD_{\xi}+{\alpha}_{2}tD_{\xi
}^{2})}.\label{t3}%
\end{equation}
We can also defined $T_{\substack{\\\phi}}$ in the non stationary case where
$\phi$ depends on time (or equivalently, as we have already seen, when $x$,
$\xi$ themselves evolve in time) and a straightfarward calculation using
eq.(\ref{phitout},\ref{markt}), show that for the case of the real line merged
with Grasmmann line i.e $N=2,$%
\begin{equation}
T_{\phi_{t}}=(\phi_{t}\otimes id_{\Omega})\circ\Delta_{\Omega}=[e^{-\lambda
t\partial/\partial a}\otimes(1-\widetilde{\lambda}t\partial/\partial
\theta)]T_{\phi_{t=0}},\label{t4}%
\end{equation}
where $T_{\phi_{t=0}}$ is as in eq.(\ref{t1}). Taking the continue limit of
the non stationary case, as we did in the case of $\phi$ , leads to%

\begin{equation}
T_{\phi_{t}}^{\infty}=e^{(-c_{1}+\lambda)tD_{x}\otimes id_{\mathcal{B}%
}+{\alpha}_{1}tD_{x}^{2}\otimes id_{\mathcal{B}}+id_{\mathcal{A}}\otimes
(c_{2}-\tilde{\lambda})tD_{\xi}^{*}}.\label{t5}%
\end{equation}
Concluding this section we must point out that we can have a left and a right
transition operators $T_{\substack{\\\phi}}^{L}$, $T_{\substack{\\\phi}}^{R},$
defined respectively by eq.(\ref{markt}) and $T_{\substack{\\\phi}%
}^{R}=(id_{\substack{\\\Omega}}\otimes\phi)\circ{\Delta}_{{\Omega}} $, and so
the above constructions have a left and right version. Also in the case where
the functional is identified with the counit i.e $\phi=\varepsilon
_{\substack{\\\Omega}}=\int$, the respective transition operators
$T_{\substack{\\\phi=\varepsilon_{\Omega}}}^{L}=T_{\substack{\\\phi
=\varepsilon_{\Omega}}}^{R}=id_{\Omega}$. and their dual probability density
$\rho_{\substack{\\\phi=\varepsilon_{_{\Omega}} }}=I_{\Omega}$, is trivial.

\section{Entanglement and statistical correlations}

We are now in a position to generalize the procedure of the previous sections
to obtain diffusion equations when the choice of the density $\rho(x,\xi)$ has
the general form of a convex combination as:%

\begin{equation}
\rho(x,\xi)=\sum_{i=1}^{m}\lambda_{i}\rho_{1}^{i}(x)\otimes\rho_{2}^{i}%
(\xi),\text{ }\lambda_{i}\geq0\text{, \ }\sum_{i=1}^{\text{ }m}\lambda
_{i}=1.\label{rgen}%
\end{equation}
>From relations (\ref{phi}) and (\ref{f}) we have that%

\begin{equation}
\phi_{\substack{\\\Omega}}=\sum_{i=1}^{m}\lambda_{i}\phi_{x}^{i}\otimes
\phi_{\substack{\\\xi}}^{i},\label{phigen}%
\end{equation}
which in the case $m=1$ and thus $\lambda_{1}=1$, $\rho(x,\xi)$ reduces to the
form of eq.(\ref{phin}) which has explicitly dealt with in the previous
chapters. The choice $m=2$ and thus $\lambda_{1}+\lambda_{2}=1$, $\rho(x,\xi)
$ reduces to the form%

\begin{equation}
\rho_{\Omega}(x,\xi)=\lambda\rho_{1}(x)\otimes I_{\mathcal{B}}+(1-\lambda
)I_{\mathcal{A}}\otimes\rho_{2}(\xi)\label{r12}%
\end{equation}
which implies that%

\begin{equation}
\phi_{\Omega}=\lambda\phi_{x}\otimes\varepsilon_{\mathcal{B}}+(1-\lambda
)\varepsilon_{\mathcal{A}}\otimes\phi_{\substack{\\\xi}}\label{phi12}%
\end{equation}
Making the choices $\rho_{1}^{i}(x)=p_{1}^{i}\delta(x-a^{i})+(1-p_{1}%
^{i})\delta(x+a^{i})$, $\rho_{2}^{i}(\xi)=p_{2}^{i}\delta(\xi-\theta
^{i})+(1-p_{2}^{i})\delta(\xi+\theta^{i})$, the general choice of
eq.(\ref{rgen}) leads to the following Taylor expanded form of $\phi_{\Omega
}^{\star n}$ :%

\begin{align*}
\phi_{\Omega}^{\star n}  & =[\sum_{i=1}^{m}\lambda_{i}(p_{1}^{i}e^{a^{i}D_{x}%
}+(1-p_{1}^{i})e^{-a^{i}D_{x}})\otimes(p_{2}^{i}e^{\theta^{i}%
D_{\substack{\\\xi}}}+(1-p_{2}^{i})e^{-\theta^{i}D_{\substack{\\\xi}}})]_{\mid
x=0,\xi=0}^{n}\\
& =[\sum_{i=1}^{m}\lambda_{i}(\varepsilon_{\substack{\\\Omega}}+2a^{i}%
(p_{1}^{i}-1/2)D_{x}\otimes\varepsilon_{\mathcal{B}}+a^{i2}/2!D_{x}^{2}%
\otimes\varepsilon_{\mathcal{B}}+...)\\
& \star(\varepsilon_{\substack{\\\Omega}}+\varepsilon_{\mathcal{A}}%
\otimes2\theta^{i}(p_{2}^{i}-1/2)D_{\substack{\\\xi}}+\varepsilon
_{\mathcal{A}}\otimes({\theta^{i})}^{2}/[2]_{q}!D_{\substack{\\\xi}%
}^{2}+...)]_{\mid x=0,\xi=0}^{n}%
\end{align*}
{We now make the substitutions}%

\begin{align}
{2a^{i}(p_{1}^{i}-1/2)}  & ={\frac{c_{1}^{i}t}n,}\text{ }{2{\theta}}%
^{i}{(p_{2}^{i}-1/2)=\frac{c_{2}^{i}t}n,}\nonumber\\
(a^{i})^{2}{/2}  & ={\frac{{\alpha}_{1}^{i}t}n,}\text{ }({\theta^{i})}%
^{2}{/[2]_{q}=\frac{{\alpha}_{2}^{i}t}n,}\label{subgen}%
\end{align}
which lead to%

\begin{align*}
\phi_{\Omega}^{\star n}  & =\{\sum_{i=1}^{m}\lambda_{i}[\varepsilon
_{\substack{\\\Omega}}+\frac1n(c_{1}^{i}tD_{x}+{\alpha}_{1}^{i}tD_{x}%
^{2})\otimes\varepsilon_{\mathcal{B}}+\frac1n\varepsilon_{\mathcal{A}}%
\otimes(c_{2}^{i}tD_{\substack{\\\xi}}+{\alpha}_{2}^{i}tD_{\substack{\\\xi
}}^{2})\\
& +\frac{t^{2}}{n^{2}}(c_{1}^{i}D_{x}+{\alpha}_{1}^{i}D_{x}^{2})(c_{2}%
^{i}D_{\substack{\\\xi}}+{\alpha}_{2}^{i}D_{\substack{\\\xi}}^{2})]\}_{\mid
x=0,\xi=0}^{n}.
\end{align*}
By implementing the limit $\lim_{n\rightarrow\infty}(1+z/n+w/n^{2})=e^{z}$ ,
it is easily obtained that%

\begin{equation}
\phi_{\Omega}^{\infty}=\exp t\{\sum_{i=1}^{m}\lambda_{i}[(c_{1}^{i}%
D_{x}+{\alpha}_{1}^{i}D_{x}^{2})\otimes\varepsilon_{\mathcal{B}}%
+\varepsilon_{\mathcal{A}}\otimes(c_{2}^{i}D_{\substack{\\\xi}}+{\alpha}%
_{2}^{i}D_{\substack{\\\xi}}^{2})]\}_{\mid x=0,\xi=0}.\label{phigeninf}%
\end{equation}
Defining the drift and diffusion terms as $L_{\mathcal{A},i}^{drift}=c_{1}%
^{i}D_{x}\otimes\varepsilon_{\mathcal{B}}$, $L_{\mathcal{B},i}^{drift}%
=\varepsilon_{\mathcal{A}}\otimes c_{2}^{i}D_{\substack{\\\xi}}$,
$L_{\mathcal{A},i}^{diff}={\alpha}_{1}^{i}D_{x}^{2}\otimes\varepsilon
_{\mathcal{B}} $, and $L_{\mathcal{B},i}^{diff}=\varepsilon_{\mathcal{A}%
}\otimes{\alpha}_{2}^{i}D_{\substack{\\\xi}}^{2}$ , eq.(\ref{phigeninf}) can
be written as%

\begin{equation}
\phi_{\Omega}^{\infty}=\exp t\{\sum_{i=1}^{m}\lambda_{i}(L_{\mathcal{A}%
,i}^{drift}+L_{\mathcal{B},i}^{drift}+L_{\mathcal{A},i}^{diff}+L_{\mathcal{B}%
,i}^{diff})\}_{\mid x=0,\xi=0}.\label{phifinal}%
\end{equation}
The form of the transition operator in this general case is shown to be%

\begin{equation}
T_{\Omega}^{\infty}=\exp t\{\sum_{i=1}^{m}\lambda_{i}[(-c_{1}^{i}D_{x}%
+{\alpha}_{1}^{i}D_{x}^{2})\otimes id_{\mathcal{B}}+id_{\mathcal{A}}%
\otimes(c_{2}^{i}D_{\substack{\\\xi}}^{*}+{\alpha}_{2}^{i}D_{\substack{\\\xi
}}^{*2})]\},\label{tfinal}%
\end{equation}
while the diffusion equation reads%

\begin{equation}
\frac{\partial\rho_{\Omega}^{\infty}}{\partial t}=\sum_{i=1}^{m}\lambda
_{i}(-L_{\mathcal{A},i}^{drift}+L_{\mathcal{B},i}^{drift}+L_{\mathcal{A}%
,i}^{diff}+L_{\mathcal{B},i}^{*diff})\rho_{\Omega}^{\infty},\label{diffinal}%
\end{equation}
where $L_{\mathcal{A},i}^{drift}=c_{1}^{i}D_{x}\otimes id_{\mathcal{B}}$,
$L_{\mathcal{B},i}^{*drift}=id_{\mathcal{A}}\otimes c_{2}^{i}%
D_{\substack{\\\xi}}^{*}$, $L_{\mathcal{A},i}^{diff}={\alpha}_{1}^{i}D_{x}%
^{2}\otimes id_{\mathcal{B}}$, and $L_{\mathcal{B},i}^{*diff}=$
$id_{\mathcal{A}}\otimes{\alpha}_{2}^{i}D_{\substack{\\\xi}}^{*2}$ . The above
equation can be rewritten as%

\begin{equation}
\frac{\partial\rho_{\Omega}^{\infty}}{\partial t}=(-c_{1}D_{x}\otimes
id_{\mathcal{B}}+{\alpha}_{1}D_{x}^{2}\otimes id_{\mathcal{B}}+c_{2}%
id_{\mathcal{A}}\otimes D_{\substack{\\\xi}}^{*}+{\alpha}_{2}id_{\mathcal{A}%
}\otimes D_{\substack{\\\xi}}^{*2})\rho_{\Omega}^{\infty}\label{newca}%
\end{equation}
where%

\begin{equation}
c_{1}=\sum_{i=1}^{m}\lambda_{i}c_{1}^{i},\ c_{2}=\sum_{i=1}^{m}\lambda
_{i}c_{2}^{i},\ \alpha_{1}=\sum_{i=1}^{m}\lambda_{i}{\alpha}_{1}^{i}%
,\ \alpha_{2}=\sum_{i=1}^{m}\lambda_{i}{\alpha}_{2}^{i}.\label{ca}%
\end{equation}

For uncorrelated walks the density function is factorized i.e $\rho
(x,\xi)=\rho_{1}(x)\otimes\rho_{2}(\xi),$ and in that case we obtain the
statistical moments%

\begin{align}
\mu_{kl}  & =<x^{k}\otimes\xi^{l}>_{\phi_{\Omega}}=\int\rho(x,\xi)x^{k}%
\otimes\xi^{l}\nonumber\\
& =\int\rho_{1}(x)x^{k}\otimes\rho_{2}(\xi)\xi^{l}=\phi_{x}(x^{k}%
)\phi_{\substack{\\\xi}}(\xi^{l})=\mu_{k}^{x}\mu_{\substack{\\l }}^{\xi
},\label{rop1}%
\end{align}
where $0\leq k<\infty$, $0\leq l\leq N-1$ and $\mu_{k}^{x}=\phi_{x}(x^{k})$,
$\mu_{\substack{\\l }}^{\xi}=\phi_{\substack{\\\xi}}(\xi^{l})$. For density
functions of the form $\rho(x,\xi)=\lambda\rho_{1}(x)\otimes I_{\mathcal{B}%
}+(1-\lambda)I_{\mathcal{A}}\otimes\rho_{2}(\xi)$ we have that%

\begin{align}
\mu_{kl}  & =<x^{k}\otimes\xi^{l}>_{\phi_{\Omega}}=\int\rho(x,\xi)x^{k}%
\otimes\xi^{l}=\lambda\phi_{x}(x^{k})+(1-\lambda)\phi_{\substack{\\\xi}%
}(\xi^{l})\nonumber\\
& =\lambda\mu_{k}^{x}+(1-\lambda)\mu_{\substack{\\l }}^{\xi}.\label{rop2}%
\end{align}
Finally, for classically correlated or seperable density functions%

\begin{equation}
\rho(x,\xi)=\sum_{i=1}^{m}\lambda_{i}\rho_{i}(x)\otimes\rho_{i}(\xi),\text{
}\lambda_{i}\geq0\text{, }\sum_{i=1}^{m}\lambda_{i}=1,\label{ro3}%
\end{equation}
we get%

\begin{equation}
\mu_{kl}=\sum_{i=1}^{m}\lambda_{i}\phi_{x}^{i}(x^{k})\phi_{\substack{\\\xi
}}^{i}(\xi^{l})=\sum_{i=1}^{m}\lambda_{i}\mu_{k}^{xi}\mu_{\substack{\\l
}}^{\xi i}.\label{rop3}%
\end{equation}

We can now procceed to obtain a differential equation for the moments. By
evaluating $\frac\partial{\partial t}<x^{k}>_{\phi_{\Omega}^{\infty}}$,
$\frac\partial{\partial t}<\xi^{l}>_{\phi_{\Omega}^{\infty}}$ and
$\frac\partial{\partial t}<x^{k}\otimes\xi^{l}>_{\phi_{\Omega}^{\infty}}$, we
obtain the equations%

\begin{align}
\frac\partial{\partial t}\mu_{k,l}(t)  & =\frac\partial{\partial t}%
<x^{k}\otimes\xi^{l}>_{\phi_{\Omega}^{\infty}}=\int(x^{k}\otimes\xi^{l}%
)(\frac\partial{\partial t}\rho_{\Omega}^{\infty})\nonumber\\
& =\sum_{i=1}^{m}\lambda_{i}[\int(x^{k}\otimes\xi^{l})(-L_{\mathcal{A}%
,i}^{drift}+L_{\mathcal{B},i}^{drift}+L_{\mathcal{A},i}^{*diff}+L_{\mathcal{B}%
,i}^{*diff})\rho_{\Omega}^{\infty}]\nonumber\\
& =\sum_{i=1}^{m}\lambda_{i}[c_{1}^{i}k\mu_{k-1,l}(t)+{\alpha}_{1}%
^{i}k(k-1)\mu_{k-2,l}(t)\label{ropes}\\
& +c_{2}^{i}[l]_{q}\mu_{k,l-1}(t)+{\alpha}_{2}^{i}[l]_{q}[l-1]_{q}\mu
_{k,l-2}(t)]\nonumber\\
& =c_{1}k\mu_{k-1,l}(t)+{\alpha}_{1}k(k-1)\mu_{k-2,l}(t)+c_{2}[l]_{q}%
\mu_{k,l-1}(t)+{\alpha}_{2}[l]_{q}[l-1]_{q}\mu_{k,l-2}(t)],\nonumber
\end{align}
where in the above derivation integration by parts has also been used for the
q-Jackson integral involving $\xi$s \cite{exton}, and $c_{1}$, $\alpha_{1}$,
$c_{2}$, $\alpha_{2}$, are given in eq.(\ref{ca} ).

In particular this last formula can be cast in two interesting reduced forms
when we consider only $x$-type diffussion or the $\xi$-type one. For the first
case, if we introduce the colume vector of $x$-statistical moments
$\mathbf{\kappa}\equiv(\mu_{k})_{k\geq0}$ where $\mu_{k}\equiv\mu_{k,0}$, and
invoke the matrix representation of the canonical Heisenberg algebra generated
by $a$, $a^{\dagger}$ and $I$ , where the number operator is $N=a^{\dagger}a$,
we can express the dynamics of the moments by means of the relation%

\begin{equation}
\frac\partial{\partial t}\mathbf{\kappa}(t)=(c_{1}a\sqrt{N}+\alpha_{1}%
a\sqrt{N}a\sqrt{N})\mathbf{\kappa.}\label{ropes1}%
\end{equation}
Similarly for the $\xi$-type case we have that
\begin{equation}
\frac\partial{\partial t}\mathbf{\nu}=(c_{2}a_{q}\sqrt{[N_{q}]_{q}}+\alpha
_{2}a_{q}\sqrt{[N_{q}]_{q}}a_{q}\sqrt{[N_{q}]_{q}})\mathbf{\nu},\label{ropes2}%
\end{equation}
where $c_{1}$, $\alpha_{1}$, and $c_{2}$, $\alpha_{2}$, are given in
eq.(\ref{ca} ), and the column vector of the $\xi$- statistical moments
$\mathbf{\nu}\equiv(\nu)_{l=0}^{N-1}=(\mu_{0,l})_{l=0}^{N-1}$ , is introduced.
The following operators
\[
a_{q}=\left(
\begin{array}
[c]{ccccccc}%
0 & \sqrt{[1]_{q}} & 0 & 0 & 0 & 0 & ...\\
. & 0 & \sqrt{[2]_{q}} & 0 & 0 & 0 & ...\\
. & . & . & . & . & . & ...\\
. & . & . & . & . & 0 & \sqrt{[N-1]_{q}}\\
. & . & . & . & . & . & 0
\end{array}
\right)  \text{,\ \thinspace}N_{q}=-\left(
\begin{array}
[c]{ccccccc}%
0 & 0 & 0 & 0 & 0 & 0 & ...\\
. & 1 & 0 & 0 & 0 & 0 & ...\\
. & . & . & . & . & . & ...\\
. & . & . & . & . & N-2 & 0\\
. & . & . & . & . & . & N-1
\end{array}
\right)  ,
\]
together with the conjugate $a_{q}^{\dagger}\ $of $a_{q},$ form a truncated
$N$-dimensional matrix representation of the so called $q$-canonical algebra
for $q=e^{i2\pi/N}$\cite{de}.

Closing we point out that eq. (\ref{ropes} ) can also be cast in the following form%

\[
\frac\partial{\partial t}\mathbf{\mu=}(\mathbf{1}_{N}\otimes L_{x}%
+L_{\substack{\\\xi}}\otimes\mathbf{1}_{\infty})\mathbf{\mu}
\]
where $\mathbf{1}_{N}$ \ is a $N\times N$ unit matrix, $\mathbf{1}_{\infty}$
is an infinite dimentional unit matrix, $L_{x}=c_{1}a\sqrt{N}+\alpha_{1}%
a\sqrt{N}a\sqrt{N}$, and $L_{\substack{\\\xi}}=c_{2}a_{q}\sqrt{[N_{q}]_{q}%
}+\alpha_{2}a_{q}\sqrt{[N_{q}]_{q}}a_{q}\sqrt{[N_{q}]_{q}}.$ Also
$\mathbf{\mu}$ is a column vector given in terms of the statistical moments of
the two partial random walks viz.%

\[
\mathbf{\mu=}[\mu_{00,}\ \mu_{10,}\ \mu_{20,}\ ...|\ \mu_{01,}\ \mu_{11,}%
\ \mu_{21,}\ ...|\ ...\ |\ \mu_{0N-1,}\ \mu_{1N-1,}\ \mu_{2N-1},\ ...]^{T}.
\]

\section{Conclusion}

In this paper we have used the machinery of Hopf algebras and the algebraic
braiding or smashing between algebras, in order to construct a non trivial
extension of a 1D random walk. In the continuous limit its diffusion equation
has been constructed and a matrix representation of it has been obtained and
solved for some spacial cases. Emphasis has been put on three aspects:
firstly, the role of algebra braiding in connection with the type of non
commutativity among the steps of the random walk, secondly, the role of
entanglement (statistical correlations) among the density probability
functions of the two partial random walks with variables $x$ and $\xi\;$and
thirdly, the role of canonical and deformed Heisenberg algebra in the
equations ruling the temporal evolution of the statistical moments of the 2D
random walk.\newpage

\textbf{Acknowledgments}

One of us (I. T.) would like to thank the National Foundation of Scholarships of Greece (IKY) for financial support.


\begin{thebibliography}{99}
\bibitem{abe}E. Abe, \textit{Hopf Algebras} (CUP Cambridge 1997).

\bibitem {asw}L. Accardi, M. Sch\"{u}rman and W. von Waldenfels, Math. Z.
\textbf{198}, 451-477 (1988).

\bibitem {drabant}Y. Bespalov and B. Drabant, \textit{Cross Product Bialgebras
- Part I-II}, math/9802028, math/9904142.

\bibitem {militaru}S. Caenepeel, I. Bogdan, G. Militaru and S. Zhu,
\textit{Smash biproducts of algebras and coalgebras}, math.qa-9809063;

\bibitem {dattoli}G. Dattoli, J. Computational and Applied Math. \textbf{118},
111 (2000).

\bibitem {davies}E. D. Davies, \textit{Quantum Theory of Open System},
(Academic, New York, 1973).

\bibitem {de}D. Ellinas, Phys. Rev. A \textbf{45}, 3358-3361 (1992).

\bibitem {ellinas}D. Ellinas, J. Comp. and Applied Math. \textbf{133} 341-353 (2001).

\bibitem {elltso}D. Ellinas and I. Tsohantjis, J. of \ Non-Linear Math.
Physics Supp. \textbf{8, } 100 (2001).

\bibitem {exton}H. Exton, \textit{q-Hypergeometric Functions and Applications,
}(Horwood, Chichester 1983).

\bibitem {qappell}P. Feinsilver and R. Schott, J. Theor. Prob. \textbf{5}, 251 (1992).

\bibitem {ffs}P. Feinsilver, U. Franz and R. Schott, J. Theor. Prob.
\textbf{10}, 797 (1997).

\bibitem {fsbook}U. Franz and R. Schott, \textit{Stochastic Processes and
Operator Calculus on Quantum Groups}, (Kluwer Academic Publishers, Dordrecht, 1999).

\bibitem {fs1}U. Franz and R. Schott, J. Phys. A: Gen . Math. \textbf{31} ,
1395 (1998).

\bibitem {franzschott}U. Franz and R. Schott, J. Math. Phys. \textbf{39}, 2748 (1998).

\bibitem {GW}N. Giri and W. von Waldenfels, Z. Wahr. Verw. Gebiete
\textbf{42}, 129-134 (1978).

\bibitem {isaev}A. P. Isaev, \textit{Paragrassmann Integral, Discrete Systems
and Quantum Groups}, q-alg/9609030.

\bibitem {Lenczewski}R. Lenczewski, Comm. Math. Phys. \textbf{154}, 127-134 (1993).

\bibitem {majidbook}S. Majid, \textit{Foundations of Quantum Groups Theory}
(Cambridge Univ. Press, 1955), ff. chapter 5.

\bibitem {majid}S. Majid, Int. J. Mod. Phys. $\mathbf{8}$ , 4521-4545 (1993).

\bibitem {majidplaza}S. Majid and M. J. Rodriguez-Plaza, J. Math. Phys.
$\mathbf{33}$, 3753-3760 (1994).

\bibitem {meyer}P. A. Meyer, \textit{Quantum Probability for Probabilists}
(Lect. Notes Math. 1538), (Springer, Berlin 1993).

\bibitem {rausch}M. Rausch de Traubenberg, Adv. Appl. Clifford Alg.
\textbf{4}, 131 (1994).

\bibitem {schurmanBook}M. Sch\"{u}rman, \textit{White Noise on Bialgebras}
(Lect. Notes Math. 1544), (Springer, Berlin 1993).

\bibitem {sLNM86}M. Sch\"{u}rman, Lect. Notes Math. \textbf{1210}, 153-157 (1986).

\bibitem {schurmanCMP91}M. Sch\"{u}rman, Comm. Math. Phys. \textbf{140},
589-615 (1991).

\bibitem {speicher}R. Speicher, Prob. Th. Rel. Fields \textbf{84}, 141-159 (1990).

\bibitem {sweedler}M. E. Sweedler, \textit{Hopf Algebras, (}Benjamin, New York 1969).

\bibitem {voiculescu}D. Voiculescu, J. Funct. Anal. \textbf{66}, 323-346 (1986).

\bibitem {W}W. von Waldenfels, Z. Wahr. Verw. Gebiete \textbf{42}, 135-140 (1978).

%
\end{thebibliography}
\end{document}